\documentstyle[aps,graphicx,times,floats,amssymb,prb]{revtex}

\newcommand{\ek}{\epsilon_{\mathbf{k}}}
\newcommand{\ekq}{\epsilon_{\mathbf{k-q}}}
\newcommand{\Ek}{E_{\mathbf{k}}}

\newcommand{\phik}{\varphi_{\mathbf{k}}}
\newcommand{\phikq}{\varphi_{{\mathbf{k}}-{\mathbf{q}}/2}}

\newcommand{\sumk}{\sum_{{\mathbf k}}}

\newcommand{\uk}{u_{\mathbf{k}}}
\newcommand{\vk}{v_{\mathbf{k}}}

\graphicspath{{$HOME/PAPERS/FIELD/FIGURES/}} 
\DeclareGraphicsExtensions{.eps,.ps}

\begin{document}
\draft

\wideabs{

  \title{Magnetic Field Effects in the Pseudogap Phase: \\ A Competing
    Energy Gap Scenario for Precursor Superconductivity }
  \author{Ying-Jer Kao$^1$, Andrew P. Iyengar$^1$, Qijin Chen$^2$ and K.
    Levin$^1$} \address{$^1$ James Franck Institute and Department of
    Physics, University of Chicago, Chicago, Illinois 60637}
  \address{$^2$National High Magnetic Field Laboratory, Tallahassee,
    Florida 32310} \date{\today} \maketitle

\maketitle
\begin{abstract}
  We study the sensitivity of $T_c$ and the pseudogap onset temperature
  , $T^*$, to low fields, $H$, using a BCS-based approach extended to
  arbitrary coupling.  We find that $T^*$ and $T_c$, which are of the
  same superconducting origin, have very different $H$ dependences.
  This is due to the pseudogap, $\Delta_{pg}$, which is present at the
  latter, but not former temperature. Our results for the coherence
  length $\xi$ fit well with existing experiments.  We predict that very
  near the insulator $\xi$ will rapidly increase.
\end{abstract}

\pacs{PACS numbers: 
74.25.Ha, 
74.60.Ec, 
74.25.-q 
\hfill \textsf{cond-mat/0103614}
}
}

One of the central questions in understanding the underdoped cuprates is
the extent to which the superconducting phase is described by BCS
theory. Recent experiments\cite{Loramreview,Krasnov1,Renner} indicate
that the pseudogap persists below $T_c$ in the underlying normal density
of states. Thus, the fermionic excitation gap $\Delta$ is to be
distinguished from the order parameter $\Delta_{sc}$. These two energy
gaps mirror a distinction between the two temperatures $T^*$(the
pseudogap onset), and $T_c$(the superconducting transition), which
behave differently as a function of hole concentration $x$ as well as of
magnetic field, $H$\cite{NMR,Zheng,Krasnov2,Renner}.  Indeed, $T^*$ and
$T_c$ are, respectively, weakly and strongly dependent on $H$ in the
well-established pseudogap regime.  Moreover, the distinction between
these temperature and energy scales has been frequently
cited\cite{NMR,Krasnov1,Loramreview} as evidence that they have
different physical origins.

In this paper we provide a counter argument to this widely stated
inference by demonstrating that these crucial magnetic field effects in
the pseudogap phase, are entirely compatible with superconductivity as
origin for \textit{both} $T^*$ and $T_c$.  Our approach is based on an
extended version of BCS theory, in which the attractive coupling $g$ is
contemplated to be strong enough so that pairs begin to form at a higher
temperature $T^*$ than the $T_c$ at which they Bose
condense\cite{NSR,Randeria}.  We have shown\cite{Kosztin1,Chen2,Chen3}
that as a necessary consequence $\Delta \ne \Delta_{sc}$.  Moreover, our
work has emphasized\cite{Kosztin1,Maly2,Chen1} that a (pseudo)gap in the
fermionic spectrum at $T_c$ is deleterious for superconductivity.  Thus,
as observed experimentally, as a function of decreasing $x$, $T_c$
decreases as the pseudogap or $T^*$ grows.

A calculation of the field dependence of $T_c$ (\textit{i.e.,} $H_{c2}$)
is an important problem in its own right. (i) This is the only way to
provide a precise interpretation of the ``coherence length" $\xi$, which
we demonstrate here is very different from that of BCS theory\cite{new}.
(ii) An analysis of $H_{c2}$ is tantamount to arriving at a
reformulation of the microscopically deduced Ginzburg-Landau (GL) free
energy up to quadratic terms, \textit{which must necessarily incorporate
  the presence of a non-zero pseudogap at $T_c$}.  (iii) Because the
``competing order parameter" scenario \cite{Loramreview,Laughlin} also
addresses the observation that $\Delta \ne \Delta_{sc}$, as well as the
competing $x$ dependences of $T_c$ and $T^*$, magnetic field effects may
provide a unique testing ground for distinguishing between these two
scenarios.

Indeed there is a rather close similarity in the structure of our
zero-field theory to the phenomenology deduced from thermodynamical data
by Loram \textit{et al}\cite{Loram}.  Our mean field
calculations\cite{Kosztin1,Chen2,Chen3} show that the gap equation for
$T\le T_c$ reduces to the usual BCS form, but with a new quasi-particle
dispersion
\begin{equation}
  \Ek =\sqrt{\ek ^2 + (\Delta_{sc}^2 +
    \Delta_{pg}^2)\phik^2}=\sqrt{\ek^2+\Delta^2\phik^2}.
\label{eq:disper}
\end{equation}
Here $\phik$ is associated with the pairing symmetry.  It follows from
this that the larger is $\Delta_{pg}(T_c)$, the lower the transition
temperature $T_c$.  In contrast to the work in Refs. \onlinecite{Loram}
and \onlinecite{Laughlin}, here $\Delta_{pg}(T)$, is determined
self-consistently and derives from the presence of a strong pairing
attraction.  \textit{Moreover, this pseudogap, $\Delta_{pg}$, persists
  below $T_c$}.

The conclusions of this paper are relatively straightforward and we
begin with a simple intuitive argument to address $H_{c2}$.  Consider
the Ginzburg-Landau free energy functional near $T_c$ to quadratic order
in $\Delta_{sc}$, in a finite field
\begin{equation}
  F\sim \left(\tau_0(T)+\eta^2 \left(\frac{\nabla}{i}-\frac{2e{\mathbf
          A}}{c}\right)^2\right) \left|\Delta_{sc}\right|^2.
\label{eq:GL}
\end{equation}
Here $\tau_0$ describes how the system approaches the critical point
with varying temperature, and $\eta^2$ is the stiffness against spatial
variations of the order parameter.  The mean-field behavior of $\tau_0$
near $T_c$ yields $\tau_0(T)=\bar{\tau}_0(1-T/T_c)$.  It follows that
 
\begin{equation}
  -\left.\frac{1}{T_c}\frac{dT_c}{dH}\right|_{H=0}=\frac{2\pi}{\Phi_0}
  \xi^2= \frac{2\pi}{\Phi_0}\frac{\eta^2}{\bar{\tau}_0}.
\label{eq:slope1}  
\end{equation}
where $\xi$ is the zero temperature coherence length.  A rough
extrapolation yields $H_{c2}(0) \approx \Phi_o/(2\pi\xi^2)$.  In the
small $g$ (i.e., BCS) case $\bar{\tau}_0=N(0)$, the density of states
per spin at the Fermi surface, and $\eta^2=N(0) 7 \zeta(3)/48 \pi^2
(v_F/T_c)^2$.  The squared coherence length $\xi_{BCS}^2=7 \zeta(3)/48
\pi^2({v_F}/{T_c})^2$ is determined by the stiffness $\eta^2$ with
$\bar{\tau}_0$ cancelling the density of states.

In contrast, in the strong coupling case the pseudogap modifies the
fermionic quasiparticle dispersion through a replacement of $\ek$ by
$\Ek$ and thereby suppresses $\bar{\tau}_0$.  Moreover, the stiffness
$\eta^2$ (which is relatively \emph{insensitive} to the energy scale of
the pseudogap), decreases due to the diminishing pair size.  There is,
thus, a competition between the numerator and denominator in
Eq.~(\ref{eq:slope1}). The decrease in $\bar{\tau}_0$ dominates at
sufficiently large coupling $g$, resulting in an extended flat region
followed by an eventual growth of $\xi^2$ with increasing $g$.  The
latter reflects the approach to the ideal ``boson" limit, where $T_c$ is
suppressed\cite{AlexScha} to zero at any $H \ne 0$.

Next, we provide a microscopic derivation of the parameters in
Eq.~(\ref{eq:slope1}) (and a related counterpart for $T^*$).  A central
theme in our paper is that both the field sensitivity of $T_c$ and of
$T^*$ in small $H$ can be studied through the zero-field normal state
pair propagator, or the inverse $t$-matrix $t^{-1}(Q) = 1/g + \chi (Q)
$, where $Q = ({\bf q},\Omega)$ is a four-vector.  The coefficients in
Eq.~(\ref{eq:GL}) will be shown to arise from an expansion of $t^{-1}$
in the momentum components perpendicular to the field (indexed by $i, j
= x, y$) as
\begin{equation}
 \tau_0=\frac{1}{g}+\chi({\mathbf 0},0), \, 
\eta^2=\frac{1}{2}
\left. \sqrt{  {\rm det} \left[ 
\partial_{q_i} \partial_{q_j}
\chi(Q) \right]} \, \right|_{Q=0}
, \label{eq:GLC}
\end{equation} 
where we have generalized from Eq. (\ref{eq:GL}) to include possible
anisotropy.

A proper motivation for our choice of $\chi(Q)$ is essential.  The
formalism\cite{Chen2,Chen3} in this paper combines a Green's function
decoupling scheme\cite{Kadanoff} with a generalization of the BCS ground
state wavefunction\cite{Leggett}.  This formalism allows for Fermi- and
non-Fermi-liquid (i.e., $\Delta_{pg}(T_c) \ne 0$) based
superconductivity, according to the size of $g$, with self consistently
determined chemical potential $\mu$.  In the present paper all technical
issues of this decoupling scheme can be simply by-passed, and the
results obtained are not only intuitive, but rather general.  All that
is needed here is the observation that the pair susceptibility
$\chi({\mathbf q}, \Omega) = \chi (Q) = \sum_{K} G(K) G_0
(Q-K)\phikq^2$.  Thus
\begin{eqnarray}
\lefteqn{\chi({\bf q},0)=\sumk\phikq^2\times}\nonumber \\
&&\left[\frac{1-f(\Ek)-f(\ekq)}{\Ek+\ekq}\uk^2
-\frac{f(\Ek)-f(\ekq)}{\Ek-\ekq} \vk^2 \right].   
\label{eq:xiq}
\end{eqnarray} 
Here $\uk^2$ and $\vk^2$ are the usual BCS coherence factors, and
$\varphi_{\bf k}^2= (1+(k/k_0)^2)^{-1}$, or $(\cos(k_x a)- \cos(k_y
a))^2$, for $s$- wave pairing in 3D jellium or $d$-wave pairing on a
quasi- 2D lattice, respectively.  That there is one full Green's
function ($G$) along with one bare Green's function ($G_0$), reflects
the structure of the BCS gap equation, which introduces\cite{Kadanoff}
this $\chi(Q)$ form, (with integrand proportional to the usual Gor'kov
$F$ function).  All numerical calculations in this paper are based on
Eqs.~(\ref{eq:disper}), (\ref{eq:GLC}), and (\ref{eq:xiq}), given
$\Delta_{pg} \equiv \Delta_{pg} (T_c)$.  Although here we proceed more
self consistently, our analytical scheme for computing the various
energy scales\cite{Chen1} can be by-passed, if $\Delta_{pg}$ and $T_c$
are pre-determined, \textit{e.g.,} fitted to cuprate experiments.

We next turn to $T^*$, where the Fermi liquid begins to break down; this
is associated with the onset of a resonance\cite{Janko,Maly1} in $t(Q)$,
as $g$ becomes sufficiently large.  Detailed numerics\cite{Maly1,Maly2}
based on the coupled Green's function equations show that to a good
first order approximation this resonance temperature can be deduced from
the condition $ 1/g + \chi_0 ({\mathbf 0},0) =0$, where $\chi_0$ is
given by Eq.~(\ref{eq:xiq}) with $\epsilon_k$ substituted for $\Ek$.
Indeed, quite generally, at $ T \ge T^*$, the $t$-matrix can be well
approximated by using $\chi_0 (Q)$ in place of $\chi (Q)$.

Magnetic field effects can be readily included into our formalism.  We
begin with a derivation of $T^*(H)$ to linear order in $H$.  Our
Hamiltonian consists of the field dependent kinetic energy term along
with the usual two-body pairing interaction of strength $ V_{\bf k,k'} =
g \varphi^{}_{\bf k}\varphi^*_{\bf k'}$ where we now include interactions
between pairs of non-zero net momentum.  Fluctuations of pairs with finite
momentum $({\bf q},{\bf k})$ are characterized by the correlation function
\[
D(q,k;q',k')=\left<{\mathcal T}\, b(q,k,\tau) b^\dagger(q',k',\tau')\right>,
\]
where 
$b(q,k,\tau)=e^{({\mathcal H}-\mu N)\tau} c_{q/2-k} c_{q/2+k} e^{-({\mathcal H}-\mu N)\tau}$.

Summing ladder diagrams leads to a Dyson equation $D = G_0G_0 -
G_0G_0VD$ with solution
\begin{equation}
  \mathop{\sum_{k,k'}} D\varphi_k^*\varphi^{}_{k'} =
  \frac{\widehat{\chi_0}}{1 + g \widehat{\chi_0}}
\label{eq:Dsolution}
\end{equation}
where $\widehat{\chi_0}=G_0G_0$ is the counterpart pair susceptibility
for $H \ne 0$, and the field-dependence of the bare electron propagator
$G_0$ is implemented using the semiclassical phase
approximation\cite{WHH}, elevating both $\widehat{\chi_0}$ and
$\sum_{k,k'}D\varphi_k^*\varphi_{k'}$ to integral operators
\cite{LeePayne} whose eigenvalues satisfy Eq.~(\ref{eq:Dsolution}).
This approximation allows the calculation of the eigenvalues of
$\widehat{\chi_0}$ from the zero-field pair susceptibility $\chi_0$ in
the regime $T \gg eH/mc$.  The field-induced relative phase shift
between electrons in a pair renormalizes the interaction $ V_{\bf k,k'}
$, but the effect is quadratic in $H$ and is therefore ignored here.
The pairing resonance temperature $T^*$ is defined by the appearance of
an eigenvalue $\Pi_0 = -g^{-1}$ of $\widehat{\chi_0}$, which causes
Eq.~(\ref{eq:Dsolution}) to diverge.  We define parameters $\eta^{*2}$
and $\tau_0^* = \bar{\tau}_0^*(1-T/T^*)$ analogous to those which appear
in Eq.~(\ref{eq:GLC}) and obtain
\begin{equation}
  g^{-1} + \Pi_0 = \tau_0^* + \eta^{*2}\cdot\frac{2e}{c}H = 0
\label{eq:GapEquationStar}
\end{equation}
which defines $T^*(H)$ to linear order in $H$.  The slope of $T^*(H)$ at
$H=0$ is
\begin{equation}
-\left.\frac{1}{T^*}
  \frac{dT^*}{dH}\right|_{H=0}
=\frac{\eta^{*2}}{\bar{\tau}_0^{*}}\frac{2\pi}{\Phi_0}
\label{eq:xi1}
\end{equation}
which leads to the associated ``coherence length" $\xi^{*2}
=\eta^{*2}/\bar{\tau}_0^*$.  The stiffness $\eta^{*2}$ can be explicitly
evaluated using the zero-field pair susceptibility, and
\begin{eqnarray}
\bar{\tau}_0^*  = \sumk &\phik^2& 
\left[ -f'(\ek) + \frac{d\mu}{dT} \frac{T}{\ek}\: \cdot \right. \nonumber \\
& & \left. \left(\frac{1-2f(\ek)}{2\ek} + f'(\ek)\right) 
\right]_{T=T^*}
\end{eqnarray}
Here we have included a contribution from the temperature dependence of
$\mu$.  In the weak coupling case, the chemical potential is pinned at
$E_F$, and we recover the ($s$-wave) BCS limit $\bar{\tau}_0^*=-\sumk
f'(\ek)\phik^2\approx N(0)\varphi_{k_F}^2\approx N(0)$.

\begin{figure}
\centerline{\includegraphics[angle=270,width=3in,clip]{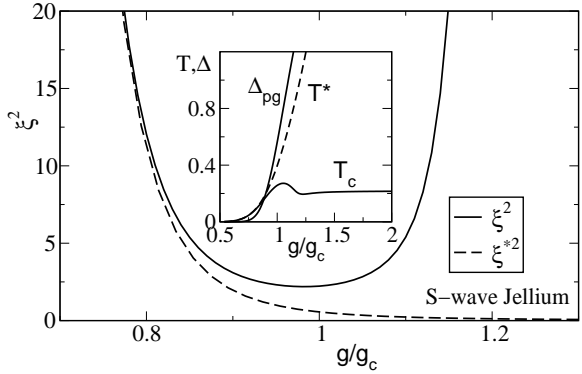}}
\caption{Coherence lengths $\xi$ (for $T_c$, solid) and $\xi^*$ 
  (for $T^*$, dashed) with variable coupling $g$ for $s$-wave jellium;
  here $g_c\equiv -4\pi/m k_0$ and units are $k_F^{-1}$.  See
  Eqs.~(\protect\ref{eq:slope1}) and (\protect\ref{eq:xi1}).  Inset
  shows the zero field behavior of $T^*$, $T_c$, and $\Delta_{pg}$ from
  Ref.~\protect\cite{Chen1}, where $T$ and $\Delta$ are in units of
  $E_F$.}
\label{fig:k4cmp}
\end{figure}

The dashed line in Fig.~\ref{fig:k4cmp} is a plot of the slope of $T^*$
with $H$, as a function of the coupling $g$, for the case of $s$-wave
jellium.  This should be compared with the inset, from
Ref.~\onlinecite{Chen1}, which illustrates the suppression of $T_c$ by
$\Delta_{pg}$ for all $g$ in the fermionic regime; (ultimately, when
$\mu$ becomes negative, $T_c$ starts to increase again). When pseudogap
effects are weak, $T^*$ is essentially the same as $T_c$ and depends
strongly on $H$.  A stronger pairing interaction $g$ causes
$\Delta_{pg}$ to increase and the pair size to decrease; the latter
effect diminishes the stiffness and causes $T^*$ to be weakly field
dependent \cite{Yamada}.

We turn next to $T_c(H)$ and note that a solution of the coupled
equations of motion\cite{Kadanoff} (as was done in the zero field case)
appears prohibitively difficult. Nevertheless, based on the above
observations that (i) in zero field $T^*$ scales rather well with
$\Delta_{pg}$ (both theoretically\cite{Chen1} and
experimentally\cite{Loramreview},) and (ii) that $T^*$ is very weakly
field dependent in the well-established pseudogap regime, we infer that
$\Delta_{pg} \equiv \Delta_{pg}(T_c)$ is weakly $H$ dependent.  This
assumption, along with the semi-classical phase approximation for the
full Green's function, are the only essential assumptions made here.
The weak $H$ dependence in $\Delta_{pg}$ appears compatible with
experiment\cite{NMR,Krasnov2,Renner} and underlies a GL formulation
(Eq.~(\ref{eq:GL})) in which only the superconducting order parameter is
coupled to the magnetic field. In this way, for the purposes of
computing $H_{c2}$, the pseudogap enters as a relatively rigid
band structure effect, which is accounted for by introducing the
\textit{full} pair susceptibility into the standard $H_{c2}$
formalism\cite{LeePayne}.  This approach necessarily yields the correct
$H =0$ result for $T_c$.

It also leads naturally to Eq.~(\ref{eq:GLC}), from which one deduces a
rather complicated expression (not shown) for $\eta^2$ along with
\begin{equation}
 \bar{\tau}_0=-\sumk \phik^2 f'(\Ek),
\label{eq:tau0}
\end{equation}
where we have omitted contributions from the temperature dependence of
$\mu$ and $\Delta_{pg}$ near $T_c$.  Eq.~(\ref{eq:tau0}) contains the
essential physics introduced by the pseudogap.  The summation
essentially measures the $E=0$ density of states which is depleted by
the pseudogap at strong coupling, leading to $\bar{\tau}_0 \sim
e^{-\Delta_{pg}/T_c} $.  Moreover, we have shown analytically that the
neglected terms further suppress $\bar{\tau}_0$, and, therefore, do not
qualitatively change our results.

\begin{figure}
\centerline{\includegraphics[angle=270,width=3in,clip]{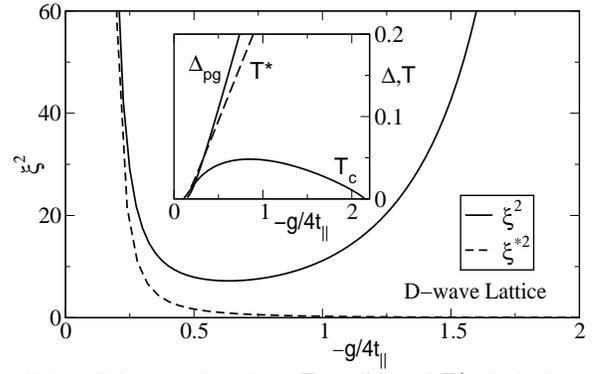}}
\caption{
  Coherence lengths at $T_c$ (solid) and $T^*$ (dashed) versus variable
  coupling $g$ for a $d$-wave lattice at density $n=0.85$. $\xi$ is in
  units of $a$.  The inset plots the zero field energy scales from
  Ref.~\protect\cite{Chen1}, and $T, \Delta$ are in units of
  $4t_{\parallel}$. }
\label{fig:dwave}
\end{figure}

Figures~\ref{fig:k4cmp} and ~\ref{fig:dwave} indicate how the
characteristic low-field slopes behave as a function of coupling $g$,
for the $s$-wave (jellium) and $d$-wave (lattice) cases, respectively.
For the latter, the bare band dispersion was presumed to be $\ek=
2t_{\parallel} (2-\cos (k_x a) - \cos (k_y a)) -2t_{\perp} (1-\cos
(k_{\perp} d)) -\mu$, where $a$ is the lattice constant in the plane and
$d$ is the distance between layers.  The insets in each figure summarize
the zero field results for $T_c$, $T^*$, and $\Delta_{pg}$, calculated
elsewhere\cite{Chen1}.  It should be noted from the insets that there is
an extended regime of coupling constants over which BCS behavior
($\Delta_{pg} =0$, and $T_c = T^*$) is obtained.  However, beyond a
``critical" coupling (corresponding roughly to where bound states of the
isolated pair occur,) the pseudogap becomes non-zero, and $T_c$ is
differentiated from $T^*$.

It is clear from the figures that the slopes of $T^*$ and $T_c$ are
identical at weak coupling and become progressively more distinct as the
coupling is increased.  The two associated stiffness parameters decrease
with coupling in a similar way, but the field dependence of $T_c$ is
enhanced by the strong suppression of $\bar{\tau}_0$ relative to
$\bar{\tau}_0^*$.  The competition between the numerator and denominator
in Eq.~(\ref{eq:slope1}) leads to a length scale $\xi^2$ which is
relatively constant over a rather wide range of moderate $g$; then at
sufficiently strong coupling $\xi^2$ begins to increase due to the
reduction in the density of states ($\bar{\tau}_0$) associated with the
growth of the pseudogap.

\begin{figure}
\centerline{\includegraphics[width=3in,clip]{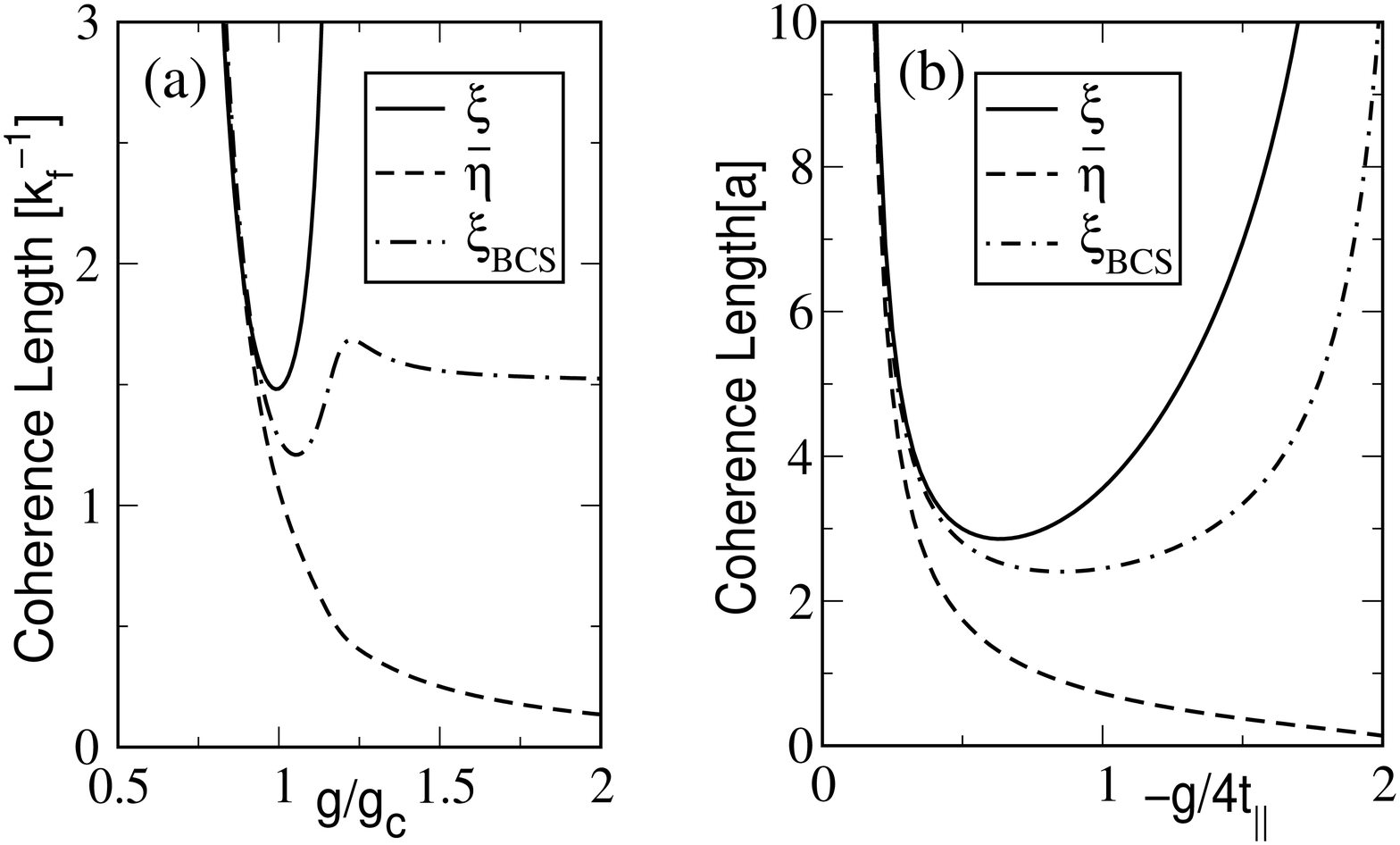}}
\caption{
  Coupling constant dependence of the key length scales: the calculated
  coherence length $\xi$ obtained from $H_{c2}$ (solid), $\bar{\eta}$
  (the normalized stiffness; dashed), and the BCS coherence length
  ($\propto v_F/T_c$; dot-dashed), for the $s$-wave jellium case (a) and
  the $d$-wave lattice case (b).  }
\label{fig:size}
\end{figure}

The key role of $\bar{\tau}_0$ in determining the squared-coherence
length $\xi^2$ highlights the fact that at strong coupling, the various
length scales of the system must be carefully distinguished even though
they are identical in BCS theory.  This is illustrated in
Figs.~\ref{fig:size}(a) and (b) for the $s$- and $d$-wave cases,
respectively, in which we compare normalized stiffness,
$\bar{\eta}=\eta/\sqrt{N(0)}$, with the BCS coherence length at $T_c$,
and $\xi$, the calculated coherence length at $T_c$.  The density of
states effect is evident in both the $s$- and $d$-wave cases, as the
upturn of $\xi$ at strong coupling contrasts sharply with the tapering
of the stiffness $\bar{\eta}$. Note that $\xi_{BCS}$ exhibits unusual
structure in Fig.~\ref{fig:size}(a) which arises from the non-monotonic
behavior of $T_c$ evident in the inset of Fig.~\ref{fig:k4cmp}. In
Fig.~\ref{fig:size}(b), the divergence of $\xi_{BCS}$ at large $g$ is
``accidental'' and derives from the vanishing of $T_c$ which arises from
pair-localization\cite{Chen1} (see Fig.~\ref{fig:dwave}, inset). BCS
relations, therefore, are very misleading when one tries to infer the
behavior of $\xi$ at strong coupling from other length scales in the
system.

\begin{figure}
\centerline{\includegraphics[angle=270,width=3in,clip]{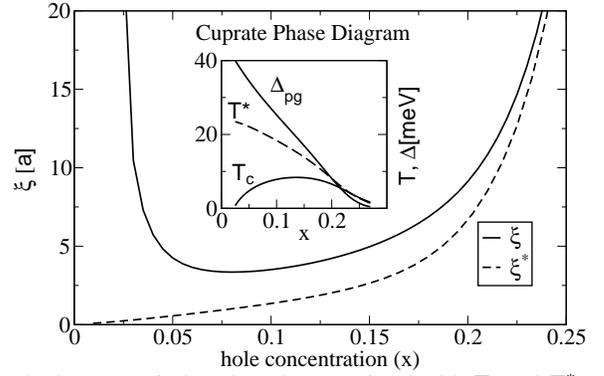}}
\caption{Magnetic length scales associated with $T_c$ and $T^*$ as a function
  of hole concentration in the cuprates. Zero field results from
  Ref.~\protect\cite{Chen2} are shown in the inset ( $\Delta_{pg}$ is a
  factor of 2 smaller than experiment due to our convention for
  $\varphi_k$ ).  }
\label{fig:doping}
\end{figure}

In order to map the coupling $g$ onto hole concentration $x$, we
introduce an $x$-dependent hopping matrix element $t_{\parallel}(x) =
t_0 x$ associated with the Mott transition\cite{Chen2}.  We presume in
the absence of more microscopic information that $g$ is $x$ independent,
leaving one free parameter in our theory $g/t_0$, chosen to optimize a
fit to the phase diagram (Fig.~\ref{fig:doping}, inset.)  One could,
alternatively, bypass all assumptions concerning Mott insulator physics,
if one instead used experimental data in place of the calculations in
the inset.  The resulting behavior of $\xi$ and $\xi^*$
(Fig.~\ref{fig:doping}) would be essentially the same.  \textit{Over
  most of the range of $x$, $\xi$ is relatively constant, as seems to be
  observed experimentally, and its magnitude is within a factor of two
  or three of experiment\cite{Wen}. As in experiment, $T^*$ is found to
  be less field sensitive in the underdoped\cite{NMR} than
  overdoped\cite{Zheng} regimes}. As the insulator is approached, $\xi$
rapidly increases \cite{Lee} while $\xi^*$ continues to decrease.  We
have thus demonstrated that the different observed field dependences of
$T^*$ and $T_c$ (for both under- and over-doped cuprates) are contained
in our theory, in which the pseudogap is associated with precursor
superconductivity.

This work was supported by NSF-MRSEC, grant No.~DMR-9808595 and by the
State of Florida (QC). We acknowledge the hospitality of the ITP, UCSB,
where the work was begun, and thank A. Carrington, B. Jank\'{o}, V.~M.
Krasnov, and Z. Te\v{s}anovi\'c for useful conversations.

\textit{Note Added.} --- After this work was completed we learned of related
experimental studies by Shibauchi \textit{et al.}\cite{Shibauchi} which show
strong similarities to our theoretical predictions.


\end{document}